\documentclass[Afour,sageh,times]{sagej}

\usepackage{moreverb,url}
\usepackage{csquotes}
\usepackage{xcolor}

\usepackage[colorlinks,bookmarksopen,bookmarksnumbered,citecolor=red,urlcolor=red]{hyperref}

\newcommand\BibTeX{{\rmfamily B\kern-.05em \textsc{i\kern-.025em b}\kern-.08em
T\kern-.1667em\lower.7ex\hbox{E}\kern-.125emX}}

\def\volumeyear{2023}

\begin{document}

\runninghead{Posada}

\title{Deeply Embedded Wages: Navigating Digital Payments in Data Work}

\author{Julian Posada}

\affiliation{Yale University}

\corrauth{Julian Posada, Yale University, American Studies, \\ P.O. Box 208369, New Haven, CT 06520-8369, United States.}

\email{julian.posada@yale.edu}

\begin{abstract}
Many of the world’s workers rely on digital platforms for their income. In Venezuela, a nation grappling with extreme inflation and where most of the workforce is self-employed, data production platforms for machine learning have emerged as a viable opportunity for many to earn a flexible income in US dollars. Platform workers are deeply interconnected within a vast network of firms and entities that act as intermediaries for wage payments in digital currencies and its subsequent conversion to the national currency, the bolivar. Past research on embeddedness has noted that being intertwined in multi-tiered socioeconomic networks of companies and individuals can offer significant rewards to social participants, while also connoting a particular set of limitations. This paper furnishes qualitative evidence regarding how this "deep embeddedness" impacts platform workers in Venezuela. Given the backdrop of a national crisis and rampant hyperinflation, the perks of receiving wages through various financial platforms include access to a more stable currency and the ability to save and invest outside the national financial system. However, relying on numerous digital and local intermediaries often diminishes income due to transaction fees. Moreover, this introduces heightened financial risks, particularly due to the unpredictable nature of cryptocurrencies as an investment. This paper evaluates the effects of the platformization of data workers' payments on their reception of wages and, ultimately, their working conditions. The over-reliance on external financial platforms erodes worker autonomy through power dynamics that lean in favor of the platforms that set the transaction rules and prices. These findings present a multifaceted perspective on deep embeddedness in platform labor, highlighting how the rewards of financial intermediation often come at a substantial cost for the workers in unstable situations, who are saddled with escalating financial risks.
\end{abstract}


\maketitle

\section{Introduction}

Platform labor has emerged as a crucial source of income for countless workers worldwide. Historically, many people in developing countries---what some term the “majority world”---have been employed in the “informal economy.” In Latin America, numerous platforms have tapped into this informal labor market, particularly in the context of the recent economic turmoil in Venezuela, which has experienced the world's highest inflation rates \citep{Long2019}. In Venezuela, high levels of unemployment, significant disruptions caused by the COVID-19 pandemic, and the existing infrastructure for remote work established during prior years of high oil revenue have combined to create a large pool of workers ideally positioned for online labor. While many of these workers are motivated by the opportunity to earn US dollars, some platforms, such as PayPal and AirTM, prefer to pay using digital currencies pegged to the US dollar. This preference compels workers to navigate transactions in virtual currencies with a range of entities, from companies to individuals.

In this article, I aim to assess the impact of platformization on data workers, particularly in terms of how it affects their receipt of wages through various online and local actors and institutions. Although platforms consider themselves intermediaries and digital infrastructures rather than direct employers \citep{Aloisi2022}, I use the term “wage” to denote their payments to workers for services rendered, even though most platforms do not recognize these workers as employees and consequently do not categorize their payments as wages. Furthermore, I build on the definition of platformization by \citet{Poell2019} to understand the “platformization of wages” as “the penetration of the infrastructures, economic processes, and governmental frameworks of platforms” into the payment and processing of online workers’ wages.

I contextualize the relationship between platform companies (referred to as firms) and actors such as workers (referred to as individuals) using the embeddedness framework from the field of economic sociology. Although various definitions of embeddedness appear in the social sciences literature \citep{Peck2013}, I use the term to describe the influence of social structures on economic activity. Here, I draw from Polanyi’s substantivism (\citeyear{Polanyi2001}), a concept that \citet{Granovetter1985} developed from a sociological standpoint. He critiqued the individualistic approach to market exchanges embraced by classical economics, focusing instead on the relationship between the various actors in a given market, including individuals and firms \citep{Uzzi1997}. Recent advancements in the area have honed in on multi-level approaches concerning the relationship between firms and individuals \citep{Brailly2016}. These advancements are especially evident in the context of platform labor, where \citet{Tubaro2021} introduced the notion of “deep embeddedness.” This concept elucidates the intricate layers of intermediation between platforms and users, as well as the rewards and constraints that these relationships engender.

I base my analysis on the wages deeply embedded in platform labor, drawing on empirical evidence collected from Venezuelan data workers between the summer of 2020 and the fall of 2022. This research focuses on data work, a specific type of platform labor. In this realm, workers contribute to generating, annotating, and verifying data for artificial intelligence (AI) via digital platforms \citep{Tubaro2020,Miceli2022}. Data work serves as a foundational task for numerous machine-learning techniques. These techniques include supervised learning, which requires labeled data for training models, and reinforcement learning, where the validation of outputs by workers helps to refine algorithms.

This overarching research project set out to analyze the configurations of outsourced data work, the power differentials inherent in the annotation process, and the impact of platform intermediation coupled with local circumstances on AI data production. The findings presented in this paper are derived from 36 semi-structured interviews with data workers. Those interviewed and surveyed conducted data work using three platforms that provide payments in digital currencies. For the sake of participant anonymity, these platforms are referred to as Clickrating, Tasksource, and Workerhub.

Clickrating and Workerhub are major players in the data production market. They allow workers to receive payments in US dollars via PayPal and also to transfer their income to cryptocurrencies via e-wallets, for example, Binance, Payeer, and Skrill. Tasksource has partnered with Airtm, a peer-to-peer e-wallet from Mexico that mediates transactions in its digital currency, AirUSD, and lets users link their accounts with several e-wallets to transfer their income to cryptocurrencies.

To evaluate the nature of deeply embedded wages in Venezuelan data work, I first provide contextual information about the economic situation in Venezuela and the rise of digital labor platforms within the country. Subsequently, I detail the network of platforms and local actors involved in processing financial payments to workers, from the point at which employers compensate workers to the conversion of digital currencies into bolivars, Venezuela’s national currency. In the paper’s final section, I conduct an analysis assessing the advantages and disadvantages of this deep embeddedness for workers, as well as the relevance of digital payments for data work more broadly.

Although access to digital payments in currencies pegged to the US dollar allows workers to bypass some of the financial restrictions of the local labor market, this has a price. While workers can save and, using cryptocurrency, invest, the varied transactional behaviors associated with digital currencies such as bitcoin \citep{Tremcinsky2022} impact their returns. These returns are further diminished by the commission fees charged by the multitude of intermediaries in this network, intermediaries ranging from financial platforms to local Venezuelan brokers that convert currencies. Additionally, the volatility of digital assets, the erosion of worker autonomy, and the absence of regulation elevate the financial risks borne by workers in their endeavors.

These findings build upon previous research on the commodification of labor in the gig economy, particularly the literature that evaluates the opportunities and risks tied to deeply embedded wage payments in platform ecosystems \citep{Wood2019,Posada2022}. The risks taken by Venezuelan workers are intensified by the local economic crisis and the absence of regulation for platform firms. Regarding embeddedness, the results indicate that not all actors benefit equally from deeply embedded networks, and those most vulnerable to market volatility ultimately shoulder a greater share of the risk.

In researching critical data studies, the findings highlight digital payments as a fundamental aspect of outsourced data work. Digital payments in the data production context have received scant research attention, despite their importance for understanding the conditions under which data for machine learning is generated when outsourced. Previous research on data work established a direct link between worker precarity and data quality \citep{Miceli2022}. The erosion of autonomy from digital payment structures not only significantly impacts our understanding of the precarization of data work and how AI companies benefit from reduced production costs but also correlates with issues such as worker alienation and declines in data quality \citep{Posada2023}. This decline in quality, in turn, affects the models trained on such data. Therefore, financial payments should be incorporated into future discussions about ethical data production practices and, more broadly, AI development.

\section{Deep Embeddedness in Platforms}

This paper examines payments in digital currencies through the lens of two theoretical frameworks: embeddedness from an economic-sociological perspective and interdisciplinary platform theory, particularly as it relates to labor and financial platforms.

The central concept that I employ here, “deep embeddedness,” was developed by \cite{Tubaro2021} and originated in the economic sociology theorization of the substantivist concept of \textit{embeddedness} \citep{Polanyi2001}. In \textit{The Great Transformation}, Polanyi used this concept in relation to the position of economic activities within institutional (e.g., governments) and noninstitutional frameworks (e.g., social relations). Polanyi simultaneously developed what \cite{Peck2013} calls a “hard” version of this concept. Peck distinguished a view of embeddedness focused on institutional and noninstitutional embeddedness (“soft” embeddedness) from a view focused on the commodification of labor, land, and money (“hard” embeddedness).

Recent research has focused on the relationship between these two definitions of embeddedness, especially in the context of platform labor \citep{Wood2019,Posada2022}. 
However, this paper emphasizes the "soft" definition primarily associated with the field of economic sociology. In his influential paper, \textit{Economic Action and Social Structure: The Problem of Embeddedness}, \cite{Granovetter1985} contended that the substantivist notion of embeddedness was overly socialized, whereas the classic economic notion of the market was under-socialized. People are not mere passive actors. Instead, they are deeply embedded in relationships and influenced by these networks. That embeddedness is also molded by social, cultural, and institutional contexts \citep{Hess2004}.

Sociological research on embeddedness is intrinsically tied to network sociology. Building on Granovetter’s foundational work, this literature has delved into the social relations influencing socioeconomic life and the social ties that affect exchanges, giving rise to a disequilibrium economy (in contrast to the equilibrium concept central to classic economics) \citep{Calnitsky2014}. A separate line of research on embeddedness stems from studies on global production networks, evolving primarily in response to the microsocial emphasis of earlier network sociology. On a macrosocial scale, scholars like \citet{Bair2008} have underscored the significance of inter-firm relationships as coordination mechanisms within expansive global value chains. In a more contemporary twist, \citet{Brailly2016} merged these approaches---inter-firm and individual-to-individual---in a multi-level network analysis that concurrently examines both the firm and individual dimensions.

In this context, \cite{Tubaro2021} introduced the concept of “deep embeddedness” in the article \textit{Disembedded or Deeply Embedded? A Multi-Level Network Analysis of Online Labour Platforms}. This term encompasses the multi-level model of embeddedness within platform labor, which situates workers inside network structures comprising firms and individuals. This concept connects the literature on economic sociology to research on digital platforms. These digital infrastructures act as intermediaries between end-users and complementors \citep{Poell2019} and function as both a multi-sided market and a firm \citep{Casilli2019}.

Although platforms have permeated various economic sectors, from entertainment to healthcare, the notion of deep embeddedness becomes especially salient in the realm of platform labor. In this context, digital intermediaries link workers with entities in need of services, connecting, for instance, drivers with passengers in the ride-hailing sector or website developers with designers in the online programming space. Notably, these platforms often position themselves primarily as technology applications rather than employment agencies, viewing their workers as users rather than employees. Consequently, these workers are predominantly classified as independent contractors who are governed by algorithms \citep{Aloisi2022} and reliant on digitized performance metrics \citep{Lehdonvirta2019} and experience heightened economic and social risks due to their line of work \citep{Tubaro2022,Posada2022}.

Although there is substantial research on platform labor that delves into the labor process \citep{Gandini2019}, the dynamics between workers and management \citep{Wood2019a}, and resistance mechanisms \citep{Englert2020}, there remains a notable gap. Specifically, the interplay between platform labor and other economic sectors, as well as the external social relations shaping this type of work, have been largely overlooked. The need for this research gap to be articulated has grown increasingly urgent with the rise of deeply embedded platforms \citep{Casilli2019c}. This need is particularly evident in the realm of data production for machine learning. Here, data security and corporate confidentiality have spurred the establishment of platform recruiters separate from data labeling platforms. Intriguingly, these entities operate in tandem, demanding that workers engage with the former to be considered for roles in the latter \citep{Schmidt2017}.

In this paper, I turn my attention to the financial platforms responsible for wage payments and transactions within this deeply embedded sector of the platform economy. A fundamental trait of digital platforms is their involvement in datafication processes, in particular, the acts of collecting, storing, and leveraging data \citep{Mejias2019}. Within the financial market, these processes are critical to the algorithmic processing of transactions \citep{Arvidsson2016}, the valuation of fictitious capital \citep{Parana2019}, and the perpetuation of the social and communicative facets of money within digitized systems \citep{Swartz2020}.

In this context, platform labor is not just the relationship between workers, employers, and firms mediating transactions. Instead, it involves a network of financial and security platforms in a complex web of actors and technologies. However, the lack of research into the roles of actors other than workers yet anchored in the same geography remains a research gap, and a central aspect of this study is addressing that gap.

\section{Research Design}

This paper examines the "platformization" of wages within data work, the segment of the gig economy to which AI developers outsource tasks such as data generation, annotation, and verification using labor platforms. The data presented here derives from a broader project centering on data work in Venezuela and encompasses three platforms operating within the country: Clickrating, Tasksource, and Workerhub. The following table outlines the characteristics of each platform.

\begin{table}[ht]
\centering
\caption{Primary Studied platforms}
\begin{tabular}{ccc}
\hline \cline{1-3}
\textbf{Work Platform} & \textbf{Payment Platform} & \textbf{Currency} \\
\cline{1-3}
Clickrating & Paypal & USD \\
Tasksource & AirTM & AirUSD \\
Workerhub & PayPal & USD \\
\cline{1-3} \hline
\end{tabular}
\end{table}

Venezuela presents a unique and compelling context for studying data work, not only because it ranks second to the US in terms of web traffic to worker sites \citep{Posada2022b}, but also due to its intricate socioeconomic landscape. The nation's reliance on oil became a vulnerability when global prices plummeted in 2014. This oil dependence, combined with an authoritarian government, corruption, and external sanctions, led to soaring inflation rates, shortages of goods and services, and constricted access to welfare. In 2018, the inflation rate reached a staggering 130,000\%, as per the latest official data from the country's federal government \citep{Long2019}. Against this backdrop, along with the existing government-sponsored infrastructure---such as locally produced computers and subsidized electricity and internet services---Venezuela emerged as an attractive labor source for platforms such as Clickrating, Tasksource, and Workerhub.

To probe the ways that workers navigate payments on these platforms, I conducted semi-structured interviews with 38 workers. Engaging with this hidden population presented challenges: Platforms often withhold details about their workers, keeping data on their numbers and demographics confidential \citep{Tubaro2020b}. As such, I reached out to potential interviewees through open groups on social media and messaging platforms, including Facebook, Discord, Reddit, and WhatsApp. I further refined this approach by integrating opportunity sampling with snowball sampling. After interviewing a set of workers identified through social media, I asked whether they could introduce me to colleagues affiliated with the three specified platforms. This strategy culminated in a total of 38 primary interviews conducted between February and August 2021. Subsequently, in November 2022, I revisited the study by holding follow-up interviews with four of the previously interviewed workers. I chose these specific individuals due to their insights into the digital currency market and its prevalence among their peers in Venezuela.

Although the convenience and snowball sampling methods employed helped bypass the secrecy of the platforms, they also introduced certain limitations. Being non-probabilistic in nature, these methods hinder generalizability and reproducibility, challenges inherent to researching a clandestine population. Furthermore, my personal background, including my position as a scholar, has likely shaped the research outcomes. Growing up in neighboring Colombia, I communicated with participants in Spanish, our shared native language. Notably, my upbringing in a multi-racial, working-class family from a similar geographical and cultural region as the participants informed my perspective. Meanwhile, my academic grounding in sociology and information studies substantially influenced my research approach. However, this also means that my current socio-economic position---a function of my role at a private US university---affords me certain privileges not shared by the study participants. These elements of positionality have undoubtedly impacted the results of this research and contributed to my rapport with the participants.

\section{Findings and Discussion}

\subsection{The Economic Crisis in Venezuela}

Venezuela emerged as a hub for digital labor outsourcing in part due to its economic crisis. According to the Organization of Petroleum Exporting Countries (OPEC), of which Venezuela is a member, the nation boasts the largest proven oil reserves among member countries, standing at 303.47 million barrels as of 2021. With the commodity as its primary export, accounting for 99\% of its export earnings \citep{Cheatham2021}, Venezuela has leaned heavily on this revenue source for decades. However, when oil prices took a nosedive in mid-2014, the economy took a significant hit. This economic downturn, exacerbated by political instability, international sanctions, government authoritarianism, and corruption, plunged the country into an ongoing economic crisis.

Despite the scarce comprehensive local studies on the living conditions of Venezuelans, especially after the COVID-19 outbreak, the Andrés Bello Catholic University undertook a notable quantitative investigation, a study spanning 22 states and encompassing 17,402 households \citep{InstitutoDeInvestigacionesEconomicasYSociales2021}. Published in September 2021 and based on data gathered between February and April of that year, the research presents a comparative analysis building upon socio-economic statistics collected earlier. The findings suggest that Venezuela witnessed a “loss of institutionality” characterized by diminished government welfare and a contentious new constitution that sidelined opposition parties. In 2021, the inflation rate soared to 2,348\%, earning the dubious distinction of being the world's highest, with the nation's gross domestic product having plummeted by 80\% since 2013. On the employment front, the study revealed a 51\% unemployment rate and a paltry minimum wage of below US\$20 per month. Of those employed, a significant 51.7\% operated as freelancers, marking a considerable increase from 30.6\% in 2014. Distressingly, the poverty rate surged from 32.6\% in 2013 to 94.5\% in 2021, with extreme poverty leaping from 9.3\% to a staggering 76.6\% during the same period.

The deep-seated repercussions of the economic crisis have significantly reshaped employment patterns in the country. Olivia, a 26-year-old Tasksource and Workerhub worker from Guanta, a coastal city in the state of Anzoátegui, shared her personal experience of the crisis. Guanta pertains to the metropolitan zone of the state capital of Barcelona and is situated close to oil refineries and shipping ports. During our interview, Olivia recounted her time as a petroleum engineering graduate who once served at the government-owned national oil enterprise, Petroleum of Venezuela (PDVSA). Her role was pivotal, analyzing oil quality in a laboratory before it could be refined and eventually exported. Many members of her family, including her father and aunt, also had affiliations with the company. However, the onset of the COVID-19 pandemic led to a wave of layoffs, impacting Olivia and her relatives. In the wake of job losses, many turned to the informal market to sustain themselves. Olivia detailed how several relatives launched food ventures and hair and beauty services. Meanwhile, Olivia had dabbled with data production platforms in 2016, during her time at university. Following her layoff from the oil company, she pivoted to platform work as her primary occupation.

Beyond its repercussions for employment, the economic downturn has also left a mark on the availability and affordability of goods and services. Workers painted a grim picture of the initial years of the crisis, marked by acute shortages and interminable lines at grocery stores. As the crisis deepened and hyperinflation kicked in, the bolivar's value nosedived at an alarming rate. Consequently, the US dollar swiftly emerged as the \textit{de facto} currency for commercial transactions, with the exchange rate oscillating multiple times a day. Katherine, an online worker from Urdaneta in the state of Miranda, emphasized the allure of earning in US dollars amidst this financial turmoil. Wages in bolivars, she remarked, simply were insufficient to survive:

\begin{displayquote}
When the exchange rate against the dollar spikes, all costs shoot up. If something is priced at 3 dollars, it soon becomes 4. My husband doesn't have the stability of a fixed income like I do; he's in car repairs. Thankfully, he's paid in dollars because accepting bolivars is almost laughable now. As an example, with the current rate hovering around 2,800 bolivars for 1 dollar, a person paid in the morning---say, 9:00 AM---might find themselves shortchanged by the afternoon. By 1:00 PM, that same dollar could be fetching 3,000 bolivars.
\end{displayquote}

According to Katherine and other workers, the government sets an exchange rate via the Central Bank of Venezuela, a rate that does not align with actual market values. Locally, this is termed the “parallel dollar.” In practice, workers obtain the real-time exchange rate through various social media platforms. For Katherine, sources of this critical information include Instagram accounts like EnParaleloVenezuela, DólarParaleloVenezuela, and MonitorDólar. These accounts typically post updates twice daily: once at 9:00 AM and again at 1:00 PM. Besides the US dollar, these platforms also detail exchange rates for the euro, digital currencies such as AirUSD, and prominent cryptocurrencies, primarily Bitcoin, with the latter available from a select ensemble of vendors active on social media. This phenomenon is elaborated upon in this paper’s next section.

These rates significantly impact local trading practices. Merchants adapt their prices daily in alignment with the parallel dollar's fluctuations, placing undue constraints on those whose earnings are solely in bolivars. For instance, Enrique, a Workerhub user residing in Caroní in the state of Bolivar, provided a stark portrayal of the economic landscape through his description of his household scenario: Enrique was living with his mother, aunt, and spouse, the first two retirees on a government-decreed minimum wage pension and his partner earning an equivalent wage. Only nine days after my discussion with Katherine, Enrique reported a staggering exchange rate: three million bolivars for a single dollar. This prompted the following detailed breakdown of his family's plight:

\begin{displayquote}
My spouse earns in bolivars, but the moment her salary is credited, its worth vanishes. The purchasing power collapses because everything gets exorbitantly priced. Here, the parallel dollar reigns supreme in commerce. Given that one dollar is now three million [bolivars], it's almost half the monthly minimum wage. On a salary of seven million per month, the purchasing scope is abysmal. Just to illustrate, a bag of Harina PAN [boiled maize flour], a staple here, costs nearly three million. So, with seven million, it's just about enough for a single maize flour pack—the main ingredient for arepas. Bare minimum sustenance necessitates earning around 50 dollars per week, and that's solely for food, sidelining other vital expenditures. The prices have soared astronomically, especially when juxtaposed with the parallel dollar.
\end{displayquote}

Hyperinflation in Venezuela has cornered workers, forcing them to immediately spend their bolivars due to its swift depreciation. This volatility leaves data workers with little choice: spend immediately or trade currency for more stable alternatives, whether that’s the US dollar or cryptocurrencies (e.g., bitcoin). With the plummeting employment numbers, particularly since the pandemic's onset, coupled with the bolivar's rapid erosion in value and the need to earn in dollar denominations, many interviewees have been propelled toward digital labor platforms. Such platforms offer an escape from the shackles of local economic and employment constraints, providing the possibility to earn highly desirable US dollars.

Therefore, while the economic downturn is only one facet of Venezuela's current situation, it is arguably the most pivotal reason for the country's emergence as a nexus for online work. As I've previously contended \citep{Posada2022}, the existing infrastructure, sponsored and established during the Chavez regime of the oil-rich 2000s, plays a vital role. Numerous workers operate on locally manufactured computers---whether freely distributed or available at prices significantly lower than global brands---and utilize public utilities either free, as in the case of electricity, or at subsidized rates, as in the case of internet access. However, challenges accompany these amenities, crucial as they are. Power interruptions are routine, and daytime internet speeds often lag, both of which interfere with online work processes. Still, this public infrastructure remains quintessential to the character of Venezuela's working population, providing an online gateway to global labor markets and the realms of digital currency.

\subsection{A Myriad of Platforms}

As \citet{Tubaro2021} noted, AI data production platforms, such as Clickrating, Tasksource, and Workerhub, operate within multiple layers of intermediation. These layers encompass workers, AI developers, and other stakeholders, including the clients of developers and the electronic financial platforms that are this article’s primary focus. Among these labor platforms, Clickrating and Workerhub employ PayPal for payments, with Tasksource utilizing AirTM. Both payment methods offer different advantages and feature distinct drawbacks.

On the one hand, US-based PayPal only requires an email address for workers to register, enabling them to quickly begin receiving digital payments in US dollars. For Venezuela-based users who wish to transfer money from their accounts, PayPal levies a transaction fee amounting to 0.89\% (as of June 2021). This fee escalates with increasing transaction amounts. For instance, a transaction involving US\$100 incurs a fee surpassing 5\%. Additionally, to authenticate their accounts, Venezuelan users must submit an identification document or link a credit card to PayPal. Non-verification can lead to restrictions, either from transferring funds or sometimes even receiving them.

On the other hand, AirTM, a financial platform originating in Mexico in 2015, operates through peer-to-peer transactions. Rather than enabling direct digital dollar transactions, AirTM labels its internal currency as AirUSD. Although certain clients, such as Tasksource, can deposit dollars directly, individual users have to engage in transactions with other members to acquire AirUSD. For example, a worker credited with 10 AirUSD from Tasksource would need to transact this amount with another user to receive an equivalent sum in another currency, potentially via an electronic wallet or bank account. To facilitate these transactions, AirTM charges a commission of 1.25\%. Importantly, AirUSD is exclusive to AirTM and cannot be moved externally. Like PayPal, AirTM restricts account operations until users verify their identity, employing a third-party platform called Onfido, which utilizes facial recognition to align selfies with official identification documents. Post-verification, users are permitted to trade AirUSD beyond conventional banking systems, converting them into digital wallets and other digital currencies, including cryptocurrencies.

Consequently, aside from the immediate financial challenges stemming from the crisis, notably hyperinflation, the inherent structures of these payment platforms, coupled with their commission schemes, prompt data workers to gravitate toward alternative electronic wallets. By transferring their pay to an electronic wallet, workers can avoid the steeper fees of PayPal and AirTM when conducting their day-to-day transactions. Interviews with workers indicate that the most prominent electronic wallets are Binance, Uphold, and Zelle. Similar to other platforms, these wallets demand identity verification via the submission of an identification document. For example, Binance requires a selfie that undergoes facial recognition akin to AirTM's process. These platforms empower workers by offering reduced commission rates on transfers and the ability to manage an array of digital currencies, including Bitcoin, Dogecoin, Tether, Tron, and ProxyNode.

As I have argued previously, an individual worker might manage multiple accounts on online data work platforms; conversely, several workers might jointly operate a single account \citep{Posada2021d}. This account sharing for online work implies that workers are paid via shared payment accounts (e.g., PayPal and AitTM) and that alternative electronic wallets (e.g., Binance) are also shared. Consider the case of Andrés: His cousin, a healthcare professional, received a government subsidy via AirTM. This subsidy granted not just the cousin but the entire family access to an authenticated account:

\begin{displayquote}
I use Uphold for Workerhub, but the account isn't in my name---it's in my cousin's. Within Uphold, one can manage several wallets in a combined account. My cousin assisted in setting up the wallet because I don’t have a passport. Later, I could establish an AirTM account due to a health subsidy bestowed upon my cousin as a “Healthcare Hero” during the pandemic.
\end{displayquote}

The situation Andrés describes highlights that these electronic wallets attract not only data workers but also those within their extended social circles. Additionally, many interviewees mentioned local merchants leveraging electronic wallets to conduct transactions in digital dollars and cryptocurrencies. With hyperinflation in play, e-wallets offer an accessible avenue for numerous workers to conserve and invest funds. Consider the case of Eduardo, a Clickrating worker from Coro in the state of Falcón. After acquainting himself with the appreciating value of cryptocurrency, he decided to diversify his investments using the Binance e-wallet. His maiden investment, a US\$50 sum equivalent to half a month's earnings, was mostly in Bitcoin but also in Tron and PRX. Although his investment doubled, its value waned in mid-2021 due to a series of tweets from Elon Musk in May 2021 \citep{Molla2021}. During our conversation in June 2021, Eduardo mentioned peers in online communities deliberating an investment in Dogecoin, a currency Musk had been endorsing. He was also hopeful about a resurgence in Bitcoin's value to bolster his investment. In this case, investments in cryptocurrency initially provided workers like Eduardo with an opportunity to more easily access forms of capital outside the local market. However, new risks accompanied this, particularly the volatility of cryptocurrency as an asset.

Examples like Eduardo's demonstrate how invaluable online communities can be for numerous workers. In earlier research, I detailed how workers employ these communities to bond beyond the scrutinized forums of labor platforms \citep{Posada2022}. In these spaces, workers express dissent, translate directives from English, and craft guides to address tasks. While some such communities flourish openly on platforms such as Facebook and Reddit, Eduardo's group prefers private communication channels---for example, Telegram and WhatsApp---accommodating a limited number of members. Here, admins enforce a nominal fee and identity verification for membership.

These groups primarily foster a trust network, especially for financial dealings and liaising with established brokers who handle digital currencies and bolivars. A predominant concern among interviewees was the looming threat of deceitful actors capitalizing on them. The virtual nature of their occupation, especially being remunerated in virtual currencies pegged to the US dollar, magnified this risk. Some threats have escalated to perilous levels, with Eduardo sharing unsettling anecdotes of colleagues hearing tales of extortion, where workers are coerced into paying a safety fee. Whether factual or hearsay, such narratives underscore the palpable anxiety plaguing workers who perceive neglect from the authorities, suggesting that platformization can manifest serious real-world threats.

This example also demonstrates that additional fees for data workers are not imposed solely by external platform actors from outside Venezuela. Instead, they can also emerge from within worker communities themselves. The power differentials at play, such as that implied by the authority of the group admins who require and manage these fees, as well as a security component for transactions that is present in the case of both large financial platforms(e.g., PayPal) and smaller online worker groups. Nonetheless, these smaller social media-based groups remain reliant on the messaging platforms that host them and on their governance policies. This illustrates that even in the case of local worker organizations, dependence on foreign platforms (notably from the US) persists.

In this milieu, social media groups act as gateways to trustworthy associates, primarily for financial transactions. For instance, Clickrating seldom prohibits workers from operating multiple accounts (thereby increasing income potential). An online marketplace offers these accounts for a few dollars, should workers trust the sellers. The more prevalent transactions are currency exchanges, particularly converting digital wallet funds to bolivars. Although some vendors accept online currencies, many do not, forcing most workers to exchange digital funds for bolivars when procuring local goods and services. Brokers endorsed by social media groups facilitate these conversions, ensuring that workers evade scams and other more grievous consequences.

Although two other related publications offer deeper insights into these social media groups \citep{Posada2021d,Posada2022}, it is pertinent here to highlight existing research confirming that dispersed data workers, despite their geographic separation and minimal interactions (due to platform mediation), do indeed form collective entities. This collectivization is exemplified in studies such as that published by \cite{Yin2016} about Amazon Mechanical Turk workers, and the work of \cite{Soriano2020} exploring UpWork and Onlinejobs.ph freelancers in the Philippines. The latter illuminates how these groups emerge as venues of dissent and what the authors term “entrepreneurial solidarities.” In the realm of digital payments, these groups equip workers with avenues for understanding and traversing the world of digital currencies. These attributes of data work align with the analysis of the WeChat groups of Chinese delivery workers by \cite{Yu2022a}. Here, learning, resistance narratives, and building solidarity stand out as pivotal group functions, mirroring the findings of Soriano and Cabañes.

\begin{figure*}
\centering
\includegraphics[width=\textwidth]{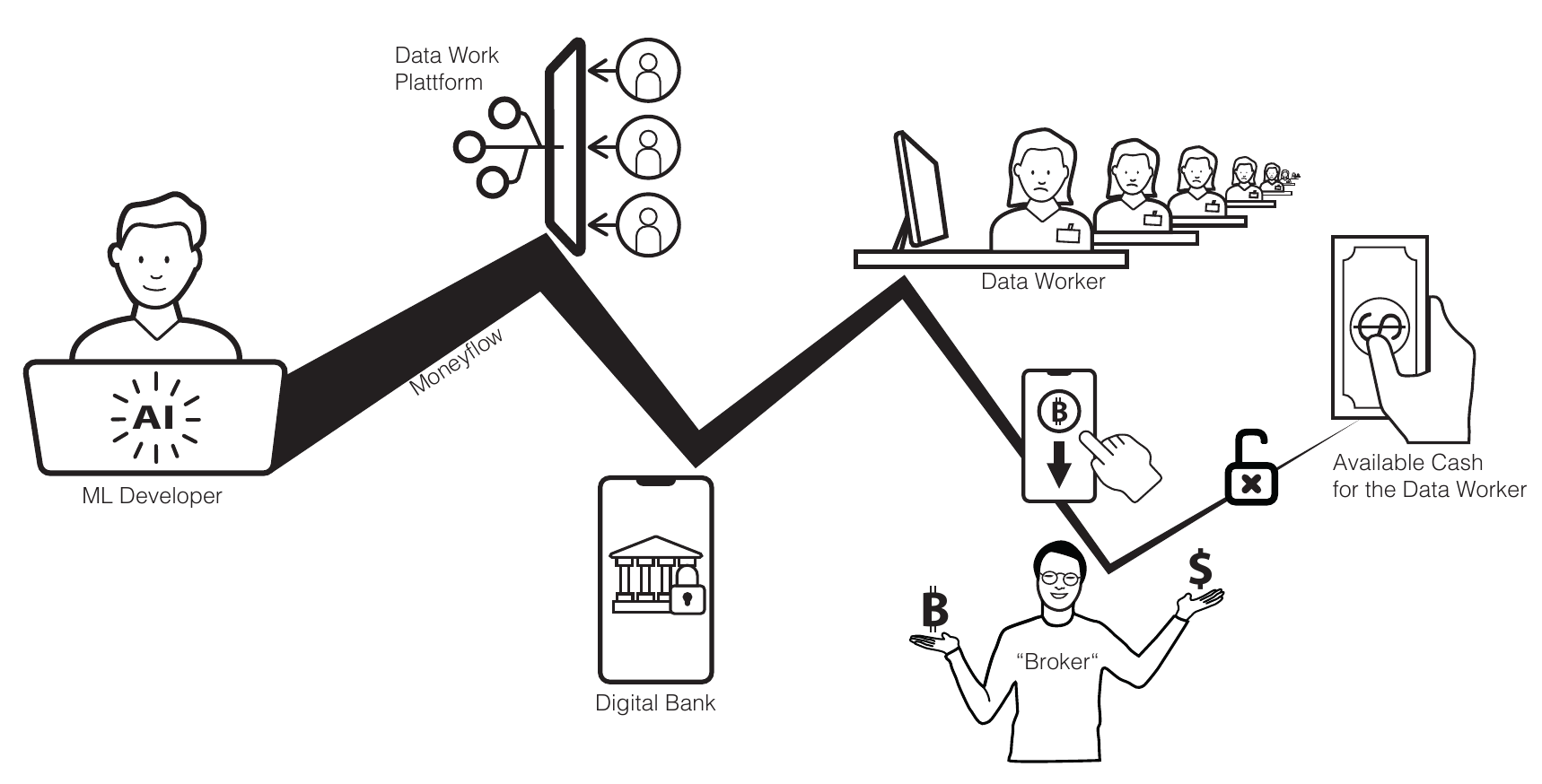}
\caption{Different actors who play a role in data work payments. Illustration by Marc Pohl.}
\end{figure*}

This section has demonstrated the ways that numerous actors participate in the transactions linked to online data work, ranging from the payment process to the expenditure of funds (refer to Figure 1). These actors include individuals (e.g., workers and brokers), communities (e.g., online worker groups and extended family members), and platforms (e.g., e-wallets and digital infrastructures designed for data production). Consequently, workers are deeply entrenched in the firm-firm, firm-individual, and individual-individual networks delineated by Tubaro. These relationships serve varied functions. Although the firm-firm and firm-individual connections revolve around the payment and handling of digital earnings, they also rely on individual-individual bonds of solidarity that are predominantly facilitated by social media platforms fostering opportunities for learning and mutual support. Up to this point, this article has delved into one layer of these networks, specifically, that associated with wages and the financial dealings of workers. The subsequent section considers the benefits and challenges stemming from this intricate embeddedness, especially with regard to wage management in the context of the ongoing Venezuelan economic turmoil.

\subsection{The Costs of Deep Embeddedness}

Despite the complicated nature of connections within the platform and between the individual actors involved in wage transactions, participation in these networks offers several benefits to workers. One notable advantage, highlighted by \cite{Wood2019}, pertains to the global online labor market that allows companies to bypass the local labor market's geographical constraints. Remembering that online platforms offer Venezuelans the opportunity to earn in US dollars, a currency considerably more stable and valuable than the bolivar, it is worth also noting that during the timeframe of this research, the consistent availability of tasks on these platforms stood in stark contrast to the country's high unemployment rates \citep{InstitutoDeInvestigacionesEconomicasYSociales2021}. Although workers noted that certain high-paying tasks were ephemeral, platforms continually offered more generalized tasks. Another advantage of sidestepping geographical limits is the flexibility to work from home, which many have emphasized as crucial. For instance, Juan, a Tasksource employee from San Fernando in the state of Apura, described working from home as follows:

\begin{displayquote}
Here, I enjoy the comfort of my home. In Colombia, my workday began at 6:00 a.m. when I was a farmer, and I'd often not return until 3:30 a.m. or even later, sometimes clocking in up to 14 hours with minimal sleep. The work was grueling. Now, I set my own pace and work on my terms and my schedule. I earn based on my effort.
\end{displayquote}

The evident advantage Juan describes here---to which he also added later the opportunity to reconnect with his family---is just one of many. In an earlier section, I explained how one benefit of using digital currencies and investing in cryptocurrencies is the ability to sidestep the bolivar's rapid depreciation, a function of the hyperinflation plaguing the nation. Therefore, for numerous Venezuelans needing access to US dollars, Colombian pesos, and other prevalent foreign currencies---especially those without foreign bank accounts---digital currencies represent the sole viable savings method. Javier, a Workerhub employee from Ciudad Ojeda, Zulia State, captured this sentiment as follows:

\begin{displayquote}
Our only savings are either on the AirTM platform or in physical dollars. We purchase the notes, confident they won't depreciate. We do maintain bank accounts (in bolivars), but only use the debit card for transactions, never for savings. Bolivars aren't viable for savings. We either maintain our money online or convert it into dollar bills.
\end{displayquote}

Javier's example highlights one issue with saving in digital currencies linked to e-wallets: users become dependent on specific services. AirUSD is only valid for AirTM transactions, while PayPal dollars are functional only when all parties utilize the platform. The network effects, a hallmark of many platforms \citep{McIntyre2017}, combine with high exit costs to anchor users---including data production platforms---to these services. While this system allows Venezuelan workers to save and invest, it also empowers platforms to bypass local financial regulations and protections, setting their own parameters, including transaction fees and authentication methods.

Another concern regarding digital currencies, especially those based on blockchain technology, is that their speculative nature often hinders them from evolving into a recognized currency \citep{Rotta2022}. For Venezuelan data workers, venturing into cryptocurrencies might produce substantial returns, as Eduardo's early investments in Bitcoin and Dogecoin suggest. However, the inherent volatility of these assets---more accurately termed speculative digital assets---introduces heightened risks into the already precarious realm of gig work. Instances like the downfall of FTX and Bitcoin's price plummet following Elon Musk's tweets underscore how, even within the cryptocurrency domain, smaller entities (i.e., workers) contribute value through their investments but also significantly bear the burden of values tumbling.

This observation aligns with the evaluation of Bitcoin by \cite{Tremcinsky2022} as a “means of exchange and storage value.” Similar dynamics play out with other currencies Venezuelans engage with, including AirTM and PayPal-hosted U.S. dollars. These currencies serve dual purposes, beyond acting as transactional media compensating data workers, they are also saving and occasionally investing instruments in a nation where hyperinflation renders the bolivar a non-viable savings option.

Furthermore, this case study substantiates Tremčinský’s concept of dual consumption spheres, echoing Jonathan Parry and Maurice Bloch’s transactional orders \citep{Parry1989}. As noted earlier, the relational dynamics of learning and community-building among data workers---also extensively documented in research by \cite{Soriano2020}---correspond to the reproduction of social orders tied to worker organization. This means that the profound integration of platforms, which facilitates an atomized workforce to tap into a distributed labor market, does not negate the emergence of worker organizations.

Furthermore, in the sphere of individual consumption, the context of digital currencies means holding them to accrue value over time, suggesting that they resemble assets more than currencies \citep{Tremcinsky2022}. Tremčinský perceives this holding incentive as a shift of power away from the state and, particularly, its monetary policies. However, in the realm of data work payments, what emerges is an inclination for powerful platform actors to define the rules of value and exchange in the digital currency space.

Thus, the ongoing platformization of wages, especially amidst financial deregulation, precipitates a decline in autonomy. Scholars have previously recognized that the tech industry operates predominantly on ideologies of individualism, self-reliance, and independence \citep{Barbrook1996,Noble2018,Benjamin2019}. The case of Venezuela's data workers exemplifies the potential pitfalls of these ideologies. Workers find minimal support, whether from institutions (in terms of welfare and market protection) or from the companies they serve within the intricately networked platform market, such as labor and financial platforms. Far from cultivating autonomous digital users, the prevailing power disparities allow firms to dictate terms and leave workers vulnerable. Although such workers do obtain some advantages, I argue that it is vital for workers to also be represented in dialogues concerning the nature of these exchanges.

This erosion of worker autonomy, especially concerning financial compensation, underscores the overarching deterioration of work environments emblematic of the gig economy. Previous investigations in the realm of computer-supported cooperative work in computer science \citep{Miceli2020} have discerned a correlation between work environments and the quality of data for machine learning. For instance, contrasting platform labor with data work in business process outsourcing setups (where a notable distinction is the deeper assimilation of workers into the data production lifecycle, including in the form of direct employment and contractual engagements), I discerned heightened levels of commitment and precision in data work. This difference was particularly prominent when juxtaposed against the surveillance-centric modalities of platform-driven data generation \citep{Miceli2022}.

Transcending the mere interplay of data quality and work conditions, prior studies have also advocated for broader discussions of critical AI ethics \citep{Ricaurte2022b}, corporate responsibilities in systems development \citep{Peters2020}, and the integration of humanistic values for the overarching betterment of all stakeholders \citep{Hadfield-Menell2019}. I argue that these interventions also require consideration of the environments in which data is cultivated, recognizing that the ethics of AI become questionable when founded on production processes that overlook core human rights associated with labor. Here, the nature of financial compensation and how workers are remunerated emerges as a pivotal determinant that actualizes the requisite conditions of ethical data production.

The expansive platform ecosystems to which the term “deep embeddedness” alludes are not limited to financial platforms but also encompass recruitment and annotation platforms that play a role in how machine learning data is generated, annotated, and validated, producing challenges around effectively gauging work conditions and, by extension, the ethics of data work. The technology industry's proclivity for opacity serves its interests, and the vagaries around work conditions spread across diverse platforms, obscuring the myriad actors and power dynamics intrinsic to data work. Consequently, future endeavors in this domain must consider preserving the merits of platform data work, such as access to stable financial resources, while concurrently addressing how this profound embeddedness potentially diminishes wages and autonomy over the long term.

\section{Conclusion}

This paper has investigated the economic ties between various firms and individual actors involved in the wage payment process for outsourced data work. Focusing on these financial transactions provides insights into the profound embeddedness of platform labor. Specifically, the paper reveals the intricate ways that workers interact with and rely upon a multitude of actors---both online and local---to facilitate payment transfers. Notably, this research underscores some constraints related to the deep embeddedness of platforms. In the Venezuelan context, although workers can tap into a global online labor market and receive payments pegged to the US dollar, the boundaries of such transactions mirror the country's unique socioeconomic dynamics and the pronounced power imbalances that advantage platforms over their workers.

Venezuela's economic and infrastructural backdrop set the stage for the rise of its data work sector. The deteriorated welfare state, coupled with soaring inflation, poverty, and unemployment rates, left a significant portion of the population dependent on freelance work, which became the primary economic activity for 51.7\% of the workforce \citep{InstitutoDeInvestigacionesEconomicasYSociales2021}. With the support of the nation's infrastructure---characterized by government-subsidized electricity and affordable computers and internet access---many Venezuelans have turned to digital platforms to bypass the constraints of the local labor and financial markets.

Workers for the three data work platforms considered here (anonymized as Clickrating, Tasksource, and Workerhub) are embedded in networks linking numerous platform companies and various local and online actors. Workers do not receive wages directly from the platform but are paid through one of two digital wallets: PayPal (for Clickrating and Workerhub) and AirTM (for Tasksource). Workers rely on local brokers, individuals who buy and sell digital currencies and bolivars, who are often recommended via social media or local networks. Workers also transfer their money to other electronic wallets, such as Binance, Uphold, and Zelle, either as a savings strategy or to invest in cryptocurrencies (e.g., Bitcoin or Dogecoin).

This deep embeddedness presents three primary challenges for workers. First, with each intermediary, workers end up with a smaller share of their earnings. Every step, whether transferring between electronic wallets or buying and selling currencies through brokers, imposes a fee that chips away at their wages. Second, storing money in digital wallets and investing in cryptocurrencies carries inherent risks. Wallet providers have their own criteria for account authentication, and workers without certain credentials (e.g., appropriate ID or bank cards) can be excluded. Additionally, fluctuations in cryptocurrency values can substantially impact a worker's personal income.

However, the most significant implications of this deep embeddedness are the consequent dependency and reduced autonomy for workers. In the realm of economic sociology, \citet{Uzzi1997} has highlighted the pitfalls of embeddedness, especially concerning vulnerability to external shocks and restricted informational access. In relation to platform involvement, \citet{Wood2019} and \citet{Tubaro2021} both emphasize the susceptibility of workers to market dynamics, given the commodified essence of platform work. The case examined in this paper demonstrates that such vulnerabilities are not confined to labor dynamics but extend to interconnected sectors including finance. More broadly, these findings indicate that deep embeddedness not only skews power dynamics between firms and individual entities but also intensifies existing inequalities within platform-worker relationships, augmenting the exit barriers for those reliant on platform labor for their primary income source. Without addressing the issues associated with digital payment, the working conditions of data workers cannot improve: ethical development of data production requires ethical payment practices. As evidenced by the literature on deep embeddedness and the examples discussed here, platform transactions are all interconnected and interdependent and their impact on the livelihoods of workers remains profound.

\begin{acks}
I want to especially acknowledge the dedication of the data worker participants who shared their knowledge and experience with me, enabling the development of these ideas. This paper was developed thanks to the feedback received in the course of several workshops and conference sessions. These included the Radical Platform Governance Workshop, the Platform Governance Research Network Conference, and the Association of Internet Researchers Annual Meeting. I also want to extend special thanks to the other authors of the special issue on Big Data \& AI in Latin America, our editor, Rafael Grohmann, and the anonymous reviewers for their feedback. Finally, I want to thank Alexander Sarra-Davis and T. Englefield for their editorial assistance and Marc Pohl for designing the illustration.
\end{acks}

\begin{funding}
This research was funded by the International Development Research Centre of Canada, the Schwartz Reisman Institute, and the Social Science Research Council.
\end{funding}

\bibliographystyle{SageH.bst}


\end{document}